\documentclass [a4paper,12pt]{article}
\usepackage{amsmath,amssymb,graphicx,cite,delarray}
\usepackage[a4paper]{geometry}
\usepackage[bf]{caption2}
\renewcommand{\captionlabeldelim}{.}
\begin{document}
\pagestyle{plain}
%\sloppy
\begin{center}\begin{Large} INFLUENCE OF THE FINITE DEFORMATIONS CHANGING THE 
SYMMETRY OF AN INITIAL LATTICE ON A GENERATION OF ATOMS DISPLACEMENTS 
WAVES BY NON-EQUILIBRIUM ELECTRONS
\end{Large}\end{center}

\begin{center}
M.P.\,Kashchenko, N.A.\,Skorikova, V.G.\,Chashchina
\end{center}

\begin{abstract}For the model electronic spectrum in the tight-binding approximation 
it is shown that the finite homogeneous deformation essentially increases the 
quantity of pairs of electronic states which are active in generation of atoms 
displacement waves. This conclusion gives additional possibilities for the 
explanation of features of the process of martensitic transformation.
\end{abstract}

The process of generation of atomic displacement waves is energized by 
stimulated emission of phonons during transitions of the non-equilibrium 
3d-electrons between the inversely occupied states. The microscopic 
theory of generation of waves controlling the martensite 
crystal growth (without taking into account deformation of a lattice) is 
in detail stated in \cite{bib:1}. The interphase region is 
characterized by the considerable chemical-potential gradient $\nabla \mu$. 
It is considered that chemical-potential gradient $\nabla \mu$ of 
electrons is defining the non-equilibrium degree. For simplicity of analysis 
it is assumed that 
$\nabla \mu$ exists in all volume. At the analysis of generation conditions 
such assumption is comprehensible 
inasmuch as the interphase region width (on the stage of growth of crystal) 
exceeds the lattice 
parameter on two or three order \cite{bib:1}. The initial inverse of population of 
pairs of electronic 
states $\sigma_{0}$ is proportional to $\nabla \mu$.

The condition of generation is resulted in a reference view:
\begin{equation} 
 \sigma_{0} > \sigma_{th},\quad \quad  \sigma_{th} = \frac{\Gamma
 \varkappa}{\vert W \vert^{2}R_{ef}}.
\label{eq1}
\end{equation}                                         
In (\ref{eq1}) $\sigma_{th}$- threshold value of an inverse populations 
difference, which is proportional to 
the inherent attenuation $\Gamma$ of radiating electrons, as well as to the
rate of damping $\varkappa$ of 
generation phonons and inversely proportional to the square of the
matrix-element $W$ of 
electron-phonon interaction, as well as to the number of pairs of equidistant 
electronic 
states $R_{ef}$. 

The calculation of $R_{ef}$ has an important significance as the execution
(\ref{eq1}) is possible 
only for the large values of $R_{ef}$. In the quasi-momentum space s-surfaces 
separate
pairs of inversely populated electronic states, on which convert in a zero 
projections 
of group speed of electrons with quasi-momentum k on a direction of spatial 
heterogeneity \cite{bib:2}. 
The equation for the s - surfaces is specified by the condition of conversion 
in zero of the 
scalar product of velocity and chemical-potential gradient vectors
\begin{equation}
(\mathbf{v}, \nabla \mu)=0.
\label{eq2}
\end{equation}
It is interesting to calculate the $R_{ef}$ in an actual region of energies
$\Delta$ of the 
order $0,1$\,eV in a vicinity of Fermi-level $\mu$. Hence it is necessary to
find the area $\Delta S$ of 
s-surface portions located between the isoenergetic surfaces $\varepsilon = \mu
\pm \Delta$ and to take into 
account that $R_{ef} \sim \Delta S$. An important stage that provides 
qualitative consideration and 
required estimation is the case of an electron energy spectrum in the 
tight-binding 
approximation for BCC and FCC lattices \cite{bib:3}. The point is that an analytical 
form of a 
spectrum allows to immediately establish an analytical form for an electron 
velocity 
field. Knowing the field one can find the s - surfaces. 

It is clear, that usage of electronic spectrum of a non-deformed (ideal) 
lattice 
is possible for small deformations and, basically, such usage is excused at 
the description 
of threshold deformation $\le 10^{-3}$ (about an elastic limit). However the 
question about an 
evolution of $R_{ef}$ during transformation is open. Let's notice, that the 
large values of 
$R_{ef}$ are connected to electronic states having the energies 
$\varepsilon \approx
 \varepsilon_{p}$ where $\varepsilon_{p}$ is a peak
of the density of states (DOS). The deformation should lead to displacement 
of a position 
$\varepsilon_{p}$ relatively of the Fermi-level $\mu$. For the symmetric peak 
of DOS not changing the form 
during deformation it would be possible to expect that the value $R_{ef}$ 
will increase at the 
decreasing of $\vert \varepsilon_{p} - \mu \vert$, whereas the value $R_{ef}$ 
will decrease at 
the increasing of $\vert \varepsilon_{p} - \mu \vert$. 
It is easy to present a non-monotonic behavior of $R_{ef}$, when during 
deformation the peak of DOS 
is displaced from a position above (below) the Fermi-level in a position 
below (above) the Fermi-level. 
Then $R_{ef}$ achieves a maximum at $\varepsilon_{p} \approx \mu$. Let's
remind that $\gamma-\alpha$ (fcc - bcc) martensitic transformation is 
a classical example of reconstructive transformation. This transformation 
is characterized by finite 
deformation (Bein's deformation \cite{bib:4}) which onto two order exceeds the 
threshold deformation. Therefore 
it is interesting to consider the influence of deformation 
$\varepsilon \sim 0,1$. 

The threshold deformation at $\gamma-\alpha$ martensitic transformation as 
the basic components contains deformation of 
a stretching along an axis of symmetry of the fourth order initial fcc 
lattice. Therefore it is expedient 
to consider influence on $R_{ef}$ of the one-axis deformations of a stretching. 
Let's designate deformation 
value $\varepsilon_{1}$, then we can write the elementary modified electron 
energy spectrum in the tight-binding approximation: 
\begin{multline}
E({\bf k}) = E_0-4 E_1\Big{(}\cos{\eta_1}\cos{\eta_2}+{}\\
{}+\frac{2}{1+(1+\varepsilon_1)^2}
\cos{\eta_3(1+\varepsilon_1)} [\cos{\eta_1}
+\cos{\eta_2}]\Big{)},
\label{eq3}
\end{multline} 
where $E_{0}$ is the atomic energy level; $E_{1}$ is the parameter characterizing 
interaction with the first neighbors 
without strain; $\eta_{i}=\mathit{a}k_{i}/2$; $\mathit{a}$ is the lattice 
parameter; and $i = 1, 2, 3$. It is considered also, that, according 
to \cite{bib:5}, $E_{1}$ is inversely proportional to a square of distance between 
the nearest neighbours. In (\ref{eq3}) a change of 
these distances at deformation is taken into account. The direction $\mathbf{e}$ 
is collinear to axis of a stretching. 
The given case is the most simple for an interpretation of dependence $R_{ef}$ 
on deformation. Really, in the density of 
states corresponding to a spectrum, there is a unique peak. Displacement and 
change of the form of this peak during 
deformation are automatically calculated. Also all sheets of s - surfaces at 
the specified direction $\mathbf{e}$ are flat 
and their contribution in $\Delta S$ can be calculated in an actual region of 
energy.

\begin{figure}[htb]
\centering
\includegraphics[clip=true, width=0.9\textwidth]{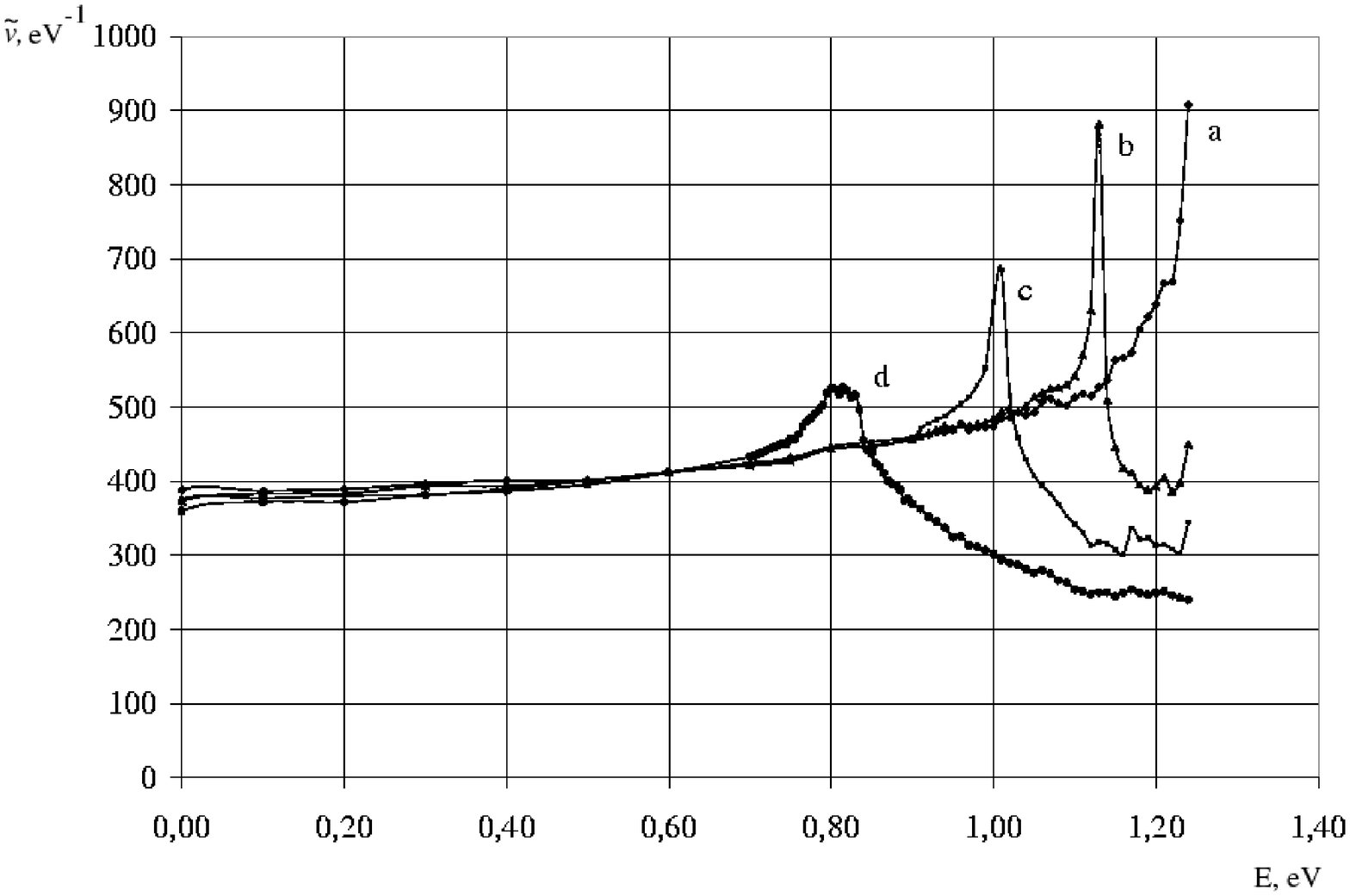}
\renewcommand{\captionlabeldelim}{.}
\caption{Evolution of peak of DOS during deformation for a spectrum (\ref{eq3}) 
at $E_{0} = 0$ and $E_{1} = 0,3125$\,eV: a) $\varepsilon_{1} = 0$; b) 
$\varepsilon_{1} = 0,1$; 
c) $\varepsilon_{1} = 0,12$; d) $\varepsilon_{1} = 0,2$.}
\label{fig1}
\end{figure}

Fig.\ref{fig1} demonstrates the change of peak of DOS during deformation.

From fig.\ref{fig1} it is visible, that: the peak of DOS is asymmetrical; 
the peak height 
decreases at the growth of $\varepsilon_{1}$; the peak of DOS localized near 
the zone top is displaced in a 
direction to the center of the zone and is broadened. Let's note, that during 
deformation Fermi-level $\mu$ also experiences the displacement to centre of 
a zone, however the rate of change of $\mu$ is 
noticeably lower, than the rate of change of $\varepsilon_{p}$.
\begin{figure}[htb]
\centering
\includegraphics[clip=true, width=0.85\textwidth]{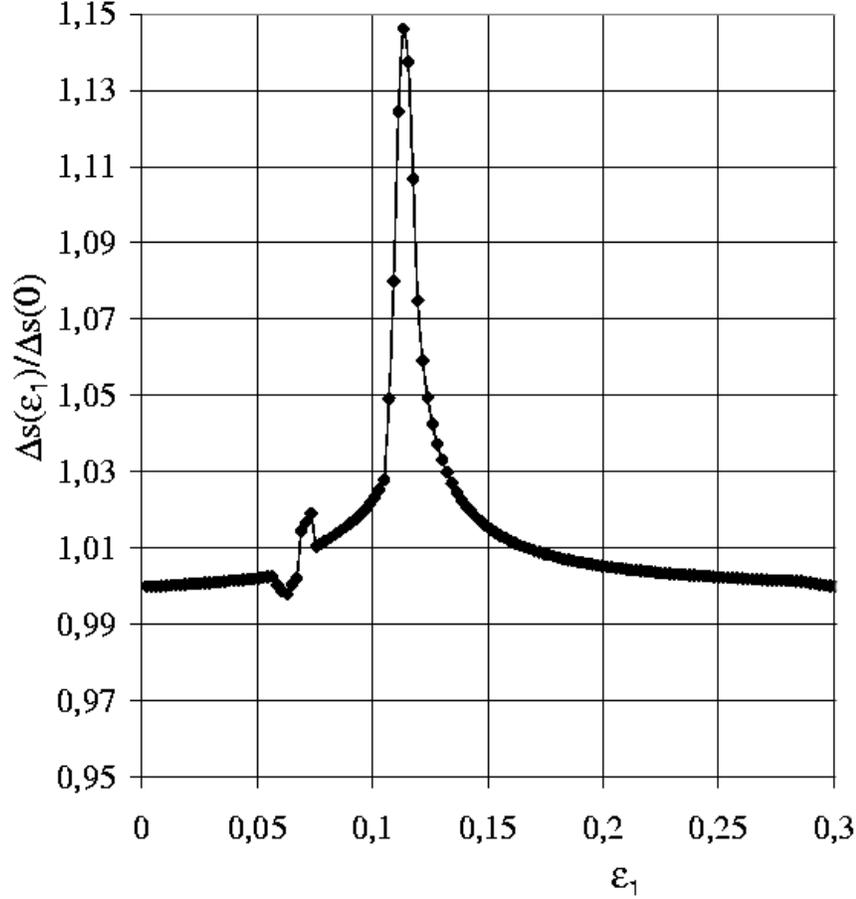}
\renewcommand{\captionlabeldelim}{.}
\caption{The dependence on deformation of dimensionless area 
$\Delta S(\varepsilon_{1})/\Delta S(0)$ for the case when $\Delta = 0,2$\,eV 
and $\mu = 0,8$\,eV.}
\label{fig2}
\end{figure}
On fig.\ref{fig2} the dependence on deformation of dimensionless area
$\Delta S(\varepsilon_{1})/\Delta S(0)$ of a s - surface portion 
is demonstrated for an energy interval $(0,6 \div 1,0)$\,eV, i.e. for the case 
when $\Delta = 0,2$\,eV and $\mu = 0,8$\,eV 
(without taking into account of the dependence of $\mu$ on $\varepsilon_{1}$).

The analysis shows that the essential increment of area 
$\Delta S(\varepsilon_{1})/\Delta S(0)$ in the strain interval $(0,06 \div 0,4)$ 
(the greatest growth of rate of $\Delta S(\varepsilon_{1})/\Delta S(0)$ is 
achieved in narrow region $0,103 < \varepsilon_{1} < 0,116$) is 
caused by the states located on square planes of Brillouin zone (orthogonal 
to $\mathbf{e}$). The area of this 
square planes increases during deformation. At $\varepsilon_{1} < 0,06$ the 
states on these planes have a weak dispersion 
and lay above an interval $(\mu - \Delta , \mu + \Delta)$. For 
$\varepsilon_{1} > 0,06$ a part of states starts to get into an energy 
interval $(\mu - \Delta , \mu + \Delta)$. Their quantity achieves of a 
maximum at $\varepsilon_{1} \approx 0,3$. Let's notices that the peak 
lays above the Fermi-level at $\varepsilon_{1} < 0,26$ $(\varepsilon_{p} - \mu > 0)$ 
and at $\varepsilon_{1} > 0,26$ lays under the Fermi-level $(\varepsilon_{p} - \mu < 0)$. 
The non-monotone behavior of $\Delta S(\varepsilon_{1})/\Delta S(0)$ as a 
whole is compounded with expected behavior. Really 
without taking into account of dependence of $\mu$ on $\varepsilon_{1}$ the 
condition $\varepsilon_{p} \approx \mu$ is realized at 
$\varepsilon_{1} \approx 0,25$ 
and $\varepsilon_{1} \approx 0,3$ corresponds to a maximum of 
$\Delta S(\varepsilon_{1})/\Delta S(0)$. Taking into account the lowering of
$\mu$ during 
deformation the condition $\varepsilon_{p} \approx \mu$ can be executed for 
$\varepsilon_{1} > 0,44$. 
The maximum $\Delta S(\varepsilon_{1})/\Delta S(0)$ corresponds 
to value $\varepsilon_{1} \approx 0,46$. Thus, taking into account of 
dependence of $\mu$ on $\varepsilon_{1}$ we obtain that the difference 
of values of $\varepsilon_{1}$ corresponding to condition 
$\varepsilon_{p} \approx \mu$ and the requirement 
$\Delta S(\varepsilon_{1})/\Delta S(0) = (\Delta S(\varepsilon_{1})/
\Delta S(0))_{max}$ 
decreases (this difference does not access in a zero, as the peak of DOS is 
not symmetric).

\begin{center} 
\textbf{Final remarks}
\end{center}

\begin{enumerate}
\item{As it was discussed in \cite{bib:1}, the interphase region width has value that 
is approximately equal $\lambda/2$, where $\lambda$ is the wave length 
corresponding to a characteristic interval $(0,1\div 1)\,\mu$m in a hypersound 
range. It is clear, that electronic states being active in wave generation 
should have a weak dispersion. It is possible to consider \cite{bib:3} that velocity
$\mathbf{v}_{\mathbf{k}}$ of such electrons is close to their velocity of 
drift vd and approximately is equal to $10^{4}$\,km/s. Then the possible 
contribution to attenuation $\Gamma$ in (\ref{eq1}) caused by occurrence of 
additional heterogeneity scale of, does not exceed (in units of frequency) $\sim
10^{12}$s$^{-1}$. Inasmuch as such $\Gamma$ is being in the consent with the 
experimental data for samples in absence of the deformation \cite{bib:1} it is natural 
to expect that deformations is influenced, mainly, on $R_{ef}$. It is clear 
that in this case a dependence $R_{ef}(\varepsilon_{1})$ gives dependence
$\sigma_{th}(\varepsilon_{1})$.}

\item{The considered example testifies about an opportunity of essential 
decrease of threshold value $\sigma_{th}$ during deformation. This conclusion 
is important  for the description of a stage of growth of the crystal of 
martensite (when the lattice experiences the significant deformation after 
loss of stability) as well as for the description of the nucleation stage of 
martensitic crystals. It is gives additional arguments in favor of the 
mechanism of rigid excitation of initial fluctuations at reconstructive 
martensitic transformations ($\sigma_{th}$ is decreasing when the amplitudes 
of fluctuations are increasing).}

\item{It is possible to assume, that most intensive martensitic transformations 
take place in alloys of such composition for which the decreasing of
$\sigma_{th}$ is typical during increasing deformation.}
\end{enumerate}

%\pagebreak
\bibliographystyle{unsrt}

\end{document}